\documentclass{ws-ijqi}
\usepackage{latexsym}
\usepackage{amsfonts}
\usepackage{bbm,dsfont}
\usepackage{graphicx}
\usepackage{amsmath}
\usepackage{color}
\newcommand{\R}{\mathbb{R}}
\newcommand{\C}{\mathbb{C}}
\newcommand{\Id}{\mathbb{I}}
\newcommand{\Sp}{\mathit{S}}
\newcommand{\B}{\mathit{B}}
\newcommand{\hil}{{H}}
\newcommand{\tr}[1]{\mathrm{Tr}\left[ {#1} \right]} 
\newcommand{\Tr}[2]{\mathrm{Tr}_{#1}\left[ {#2} \right]} 
\newcommand{\ket}[1]{\left\vert{#1}\right\rangle}

\newcommand{\ketbra}[2]{\left\vert {#1} \right\rangle\left\langle {#2} \right\vert}

\begin{document}
\markboth{Carlo Sparaciari, Matteo G. A. Paris}
{Probing qubit by qubit}
\catchline{}{}{}{}{}
\title{PROBING QUBIT BY QUBIT: PROPERTIES OF THE POVM AND
THE INFORMATION/DISTURBANCE TRADEOFF}
\author{CARLO SPARACIARI}
\address{Dipartimento di Fisica, Universit\`a degli Studi di Milano, 
I-20133 Milano, Italy\\
carlo.sparaciari@studenti.unimi.it}
\author{MATTEO G. A. PARIS}
\address{Dipartimento di Fisica, Universit\`a degli Studi di Milano, 
I-20133 Milano, Italy\\
CNISM, UdR Milano, I-20133 Milano, Italy\\
matteo.paris@fisica.unimi.it}
\maketitle
\begin{history}
\received{}
\end{history}
\begin{abstract}
We address the class of positive operator-valued measures (POVMs) for
qubit systems that are obtained by coupling the signal qubit with a probe qubit 
and then performing a projective measurement on the sole probe system. 
These POVMs, which represent the simplest class of qubit POVMs, depends 
on $3+3+2=8$ free parameters describing the initial preparation of the 
probe qubit, the Cartan representative of the unitary coupling, and the
projective measurement at the output, respectively. We analyze in some details 
the properties of the POVM matrix elements, and investigate their values for 
given ranges of the free parameters. We also analyze in details 
the tradeoff between information and disturbance for different ranges of the 
free parameters, showing, among other things, that i) {\em typical
values} of the tradeoff are close to optimality and ii) even using a maximally mixed 
probe one may achieve optimal tradeoff.
\end{abstract} 
\section{Introduction}
A common task in quantum technology is that of extracting information
about the state of a physical system without destroying the information
itself, i.e. possibly leaving part of it for another users.
This is usually accomplished through indirect measurement, i.e. coupling
the system of interest with a probe system and performing measurements
on the probe \cite{ban}. The information on the system is thus provided by the probe 
and the system is not destroyed, though its state may be changed after
the measurement. This measurement strategy may be described in terms of
the sole system, neglecting the probe, by tracing out the probe degrees 
of freedom. This procedure returns a positive operator-valued measure 
(POVM) on the Hilbert space of the system, which describes both the
statistics of the outcomes and the state reduction due to the
measurement.\cite{hel1,hel2,hol,ber,epjst}
For qubit systems the simplest class of POVMs involves
another qubit as probe and depends on $3+3+2=8$
free parameters, which describe the initial preparation 
of the probe qubit, the unitary operator coupling the two qubits, 
and the projective measurement at the output, respectively.
\par
In this paper, we address the properties of this class of POVMs 
as a function of the free parameters. In particular, in order to
obtain information about their typical values, the distribution of 
POVMs' matrix elements is analyzed for random choices of the free 
parameters in different ranges. Besides, we analyze in some
details the tradeoff between information and disturbance, showing that
typical values of the tradeoff are close to optimality and that 
even using a maximally mixed probe one may still achieve optimal tradeoff.
\par
The paper is structured as follow. In Section \ref{s1} we describe in
details the measurement scheme and the range of variation of the free 
parameters. In doing this we review the Cartan decomposition of
two-qubit unitaries and provide the characterization of the 
POVM elements, the so-called {\em effects} \cite{kc,zvsw,tuc}. 
In Section \ref{s2} we analyze the distribution of the POVM
matrix elements as a function of the free parameters. In 
Section \ref{s3} the quantification of information and 
disturbance is briefly reviewed and the corresponding 
distribution of fidelities is studied as a function of the free
parameters. Section \ref{con} closes 
the paper with some concluding remarks.
\section{The measurement scheme}\label{s1}
\begin{figure}[h!]
\centering
\includegraphics[width=0.5\textwidth]{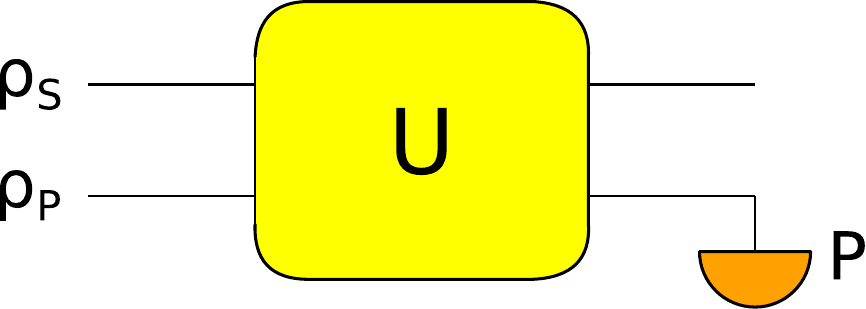}
\caption{(Color online) Schematic diagram of a general measurement
scheme exploiting a {\em probe} qubit in order to gain information 
on a {\em signal} qubit prepared in an unknown state $\rho_S$. In 
the first step the probe qubit is prepared in a known state $\rho_P$, 
then the signal and the probe are coupled by the two-qubit unitary $U$ 
and, finally, a projective measurement described by the
projection-valued measure $\{P,\Id-P\}$ is performed on the sole probe
qubit.}
\label{ms}
\end{figure}

Let us consider the following scheme of measurement, which exploits
a {\em probe} qubit in order to gain information on a {\em signal}
qubit. In the first stage the probe qubit is prepared in a known state
and then the signal and the probe are coupled by a unitary operator.
Finally, a projective measurement is performed on the sole probe 
system (see Fig.\ \ref{ms}). 
\par
The unitary operator $U$ works on $\C^2 \otimes \C^2$, and we assume its determinant
to be equal to $1$, in order to have $U \in SU(4)$.  We refer to the Hilbert
space of the system as $\hil_S$, while the Hilbert space of the probe is
$\hil_P$. 
The state of the probe $\rho_P \in
\Sp(\hil_P)$ in the Bloch representation may be written as
\begin{equation*}
\rho_P = \frac{1}{2} (\Id + \boldsymbol{r} \cdot \boldsymbol{\sigma}) 
\end{equation*}
where the Bloch vector $\boldsymbol{r} = (r_1, r_2, r_3)$ is given by
\begin{align*}
r_1 = \sqrt{2 \mu - 1} \sin \theta \cos \phi\,,\quad
r_2 = \sqrt{2 \mu - 1} \sin \theta \sin \phi\,,\quad
r_3 = \sqrt{2 \mu - 1} \cos \theta
\end{align*}
where $\mu \in [1/2,1]$ is the purity of the probe system, and
$\theta \in [0,\pi]$, $\phi \in [0,2 \pi)$. 
\par
The projective measurement, performed on the probe system, is 
described by $P = \ketbra{\xi}{\xi}$, where
$\ket{\xi} = \cos \frac{\alpha}{2} \ket{0} + e^{i \beta} \sin \frac{\alpha}{2} \ket{1}
$
and $\alpha \in [0,\pi]$ and $\beta \in [0,2 \pi)$.
Since the probe is a qubit, then the projective measurement is composed by $P$ and $\Id - P$.
\par
The unitary operator $U \in SU(4)$ depends on $15$ parameters. In order to reduce this number,
we use the Cartan decomposition, which allows us to replace $U$ with the operator
$V$, working on both system and probe, depending on just 3 parameters,
plus four local unitary operators, namely $R_1, R_2, S_1, S_2 \in SU(2)$:\cite{tuc}
\begin{equation*}
U = (R_1 \otimes R_2) V (S_1 \otimes S_2)
\end{equation*}
The operator $V$ is given by:
\begin{equation*}
V = \mathrm{exp} \left\{ -i \left[ \frac{1}{2} 
(\alpha_1 - \alpha_2)\, \Sigma_1 + \frac{1}{2} (\alpha_1 
+ \alpha_2)\, \Sigma_2 + \alpha_3\, \Sigma_3\right] \right\}
\end{equation*}
where $\Sigma_i = 1/2\ \sigma_i \otimes \sigma_i$ and the 
parameters $\alpha_i$ should satisfy the following constraints:
\begin{subequations} \label{eq:const_alpha}
\begin{align}
&-\pi \leq \alpha_1 \leq 0 \label{eq:alpha1} \\
&0 \leq \alpha_2 \leq -\alpha_1 \label{eq:alpha2} \\
&\alpha_1 + \alpha_2 \leq 2 \alpha_3 \leq 0 \label{eq:alpha3}
\end{align}
\end{subequations}
Moreover, if $\alpha_3 = 0$, then $\alpha_1 - \alpha_2 \geq -\pi$.
Clearly, the Cartan decomposition does not reduce the number 
of parameter of $U$, since each local operator depends on 3 parameters.
However, as we will see, for our purposes, the local 
operators could be neglected.
\par
The measurement scheme given above can be described by a POVM
on the Hilbert space $\hil_S$. The operators
which compose a POVM are often referred to as {\em effects}. An effect represents an apparatus
with dicotomic outcome (yes$/$no). Therefore, each effect of a POVM is connected to
a single outcome of the apparatus, and gives the probability that its
outcome occurs.\cite{lud,kra}. The effects composing this POVM are given
by the following equation (Naimark Theorem): \cite{nai}
\begin{equation}\label{effect}
\Pi = \Tr{P}{(\Id \otimes \rho_P)\ U^{\dagger}\ (\Id \otimes P)\ U}
\end{equation}
Notice that, since the PVM on the probe system is $\{ P , \Id - P \}$, then
the POVM on $\hil_S$ is composed by two effects, i.e.\ $\{ \Pi , \Id -
\Pi \}$, and it is fully characterized by the matrix elements of $\Pi$.
The Cartan decomposition of $U$ may be exploited to rewrite Eq.
(\ref{effect}) as follows
$\Pi = S_1^{\dagger}\ \Tr{P}{(\Id \otimes S_2\, \rho_P\, S_2^{\dagger})\ 
V^{\dagger}\ (\Id \otimes R_2^{\dagger} \,P \,R_2)\, V} S_1\,.
$
The local operators $R_2$ and $S_2$ are rotations in the qubit space
$\hil_P$ and may be easily eliminated by a suitable reparametrization of 
the probe state $\rho_P$ and the projector $P$. 
The rotation $S_1$ corresponds to an operation performed on the system
qubit {\em before} the measurement, and it does not affect the
properties of the POVM itself \cite{spa13}. We thus assume, without loss of
generality, to have $S_1=\Id$.
Overall, the effect $\Pi \in \B(\hil_S)$ may be written as
\begin{equation}\label{effectPi}
\Pi = \Tr{P}{(\Id \otimes \rho_P)\ V^{\dagger}\ (\Id \otimes P)\ V}
\end{equation}
In the Pauli basis we have $\Pi = a_0\ \Id + \boldsymbol{a} \cdot \boldsymbol{\sigma}$,
with $\boldsymbol{a}=(a_1, a_2, a_3)$,
where $a_0 = \frac12 \tr{\Pi}$ and $\boldsymbol{a} = \frac12 \tr{\Pi\, \boldsymbol{\sigma}}$. 
These coefficients depend on the eight free parameters $\alpha_1$, $\alpha_2$, $\alpha_3$, $\mu$,
$\theta$, $\phi$, $\alpha$ and $\beta$. The analytic expression of the coefficients of $\Pi$
is given in the \ref{a:cote}, and will be used in Section \ref{s2} to
characterize the properties of the POVM as a function of the free
parameters.
\section{Characterization of $\Pi$}\label{s2}
As mentioned above, the operator $\Pi$ fully describes the POVM and, in
turn, the measurement scheme. $\Pi$ is an effect, i.e. a bound operator, 
which is positive, and hence selfadjoint, and with eigenvalues smaller 
that 1.  Sometimes these conditions are
synthetically expressed as $0 \leq \Pi \leq \Id$ which, after straightforward 
calculations, may be shown equivalent to the following constraints:
\begin{subequations}\label{eq:constr}
\begin{align}
&0 \leq \vert a \vert \leq 1/2\\
&\vert a \vert \leq a_0 \leq 1 - \vert a \vert
\end{align}
\end{subequations}
where $\vert a \vert = \sqrt{a_1^2+a_2^2+a_3^2}$ and $(a_0,
\boldsymbol{a}) \in \R^4$. If $a_0 = \vert a \vert = 1/2$, then $\Pi$ is 
a projector, i.e. an extremal point of the set of effects.
\par
\begin{figure}[h!]
\centering
\includegraphics[width=0.32\textwidth]{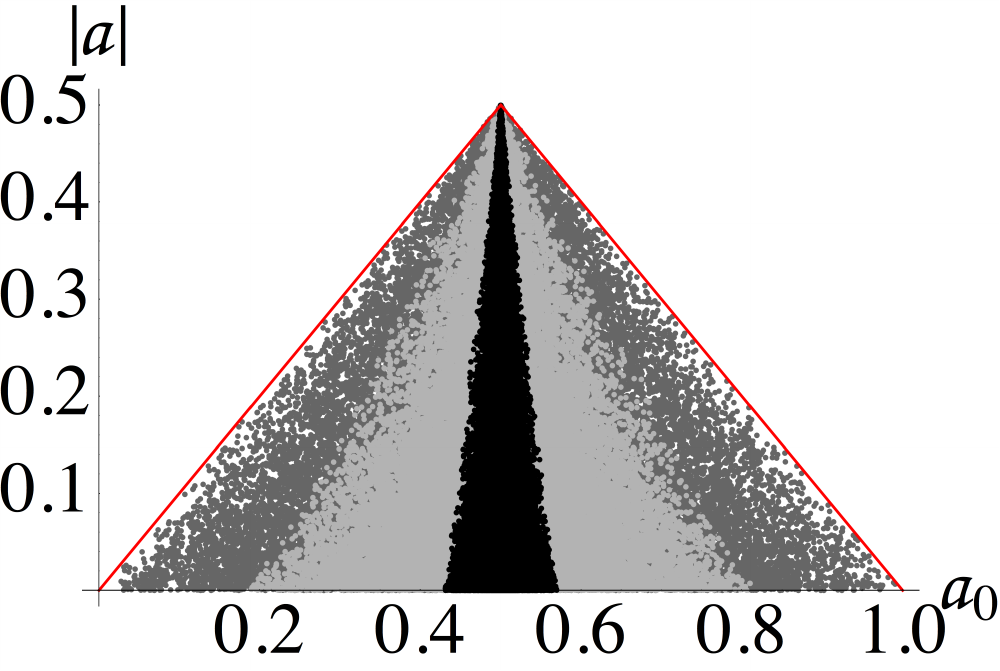}
\includegraphics[width=0.32\textwidth]{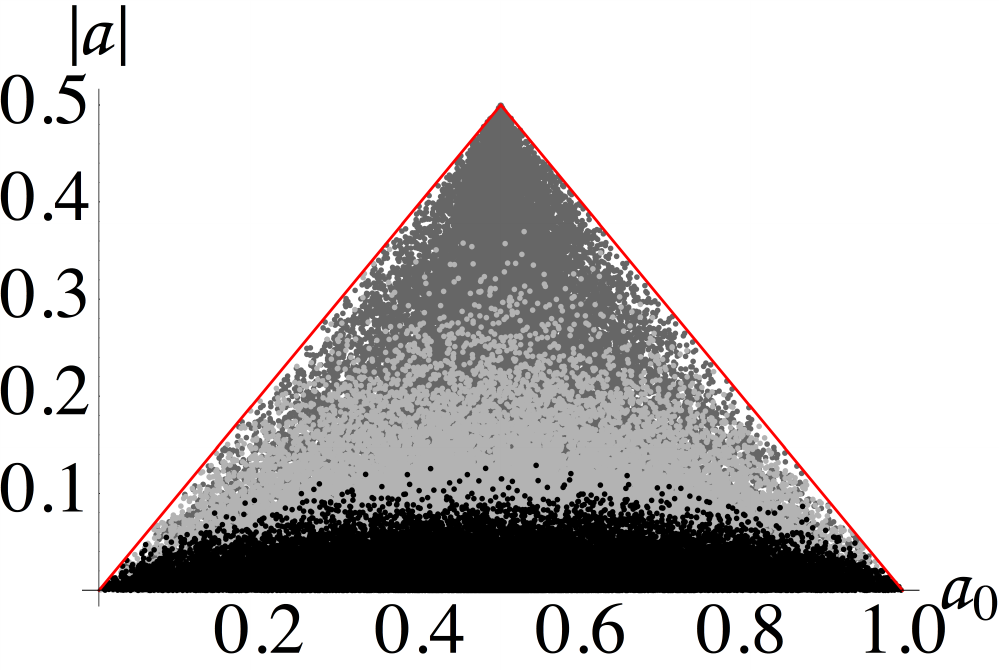}
\includegraphics[width=0.32\textwidth]{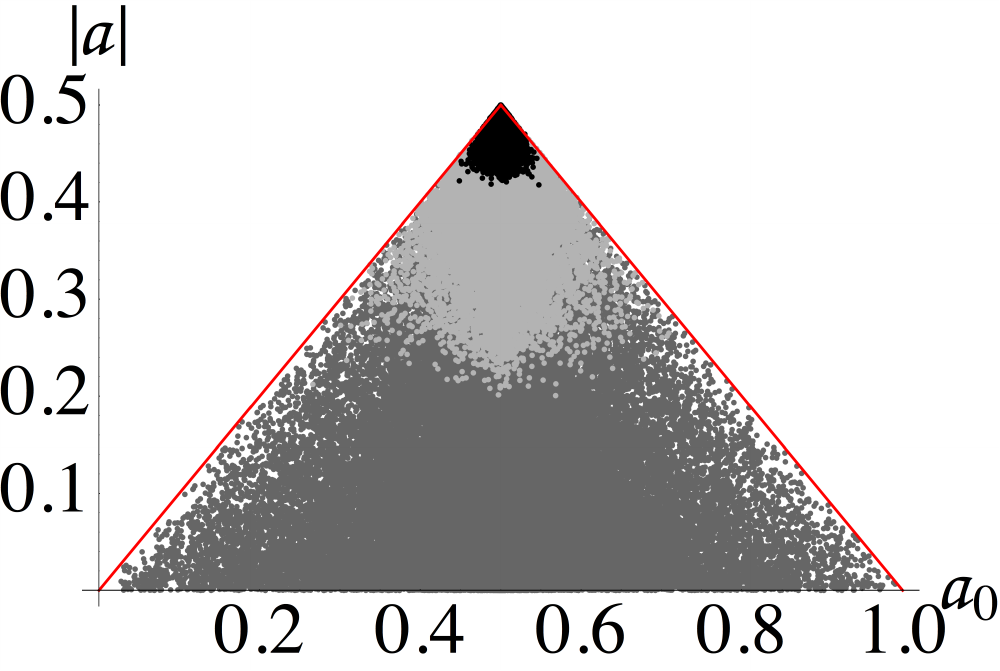}
\caption{(Color online) The distributions of $\{ a_0, \vert a \vert \}$
inside the allowed region given by Eq. (\ref{eq:constr}) 
(individuated by the red line) for different 
ranges of the free parameters. In all the plots the medium gray points 
correspond to the POVMs obtained with all the free parameters chosen at
random in their whole ranges of variation.
The light gray points and
the black ones corresponds to POVMs obtained choosing the free
parameters at random in restricted ranges. In the left panel, we show 
the POVMs corresponding to different ranges for $\mu$: light gray points 
are for $\mu \in [0.5, 0.7]$, while the black ones corresponds to 
$\mu \in[0.5,0.51]$. The center panel describes both the situations in which
the range of all the parameters $\alpha_i$ tends to 0
and in which the range of $\alpha_1$ tends to $-\pi$, the range of
$\alpha_2$ tends to $\pi$ and the one of $\alpha_3$ tends to 0 (see text).
The right panel shows the case in which the range of
$\alpha_1$ tends to $-\pi$, the range of $\alpha_2$ tends to 0 and
the one of $\alpha_3$ tends to $-\pi/2$.}
\label{f2:coeff}
\end{figure}

We now study in some details the distribution of the parameters
$a_0$ and $|a|$ within the physical region determined by Eq. (\ref{eq:constr}). 
First of all, we check whether, taking at random the values of the
free parameters in their whole ranges, we obtain a uniform distribution
in the physically allowed region. This is indeed the case, as it can be seen by looking
at the medium gray points in the three panels of Fig.\ \ref{f2:coeff}.
\par
Let us now analyze how the purity $\mu$ of the probe system affects the
properties of the POVMs: in the left panel of Fig. \ref{f2:coeff} light 
gray points are obtained by selecting $\mu$ in the range $[0.5, 0.7]$, 
while the black ones are obtained using a range $[0.5,0.51]$. 
As it is apparent from the plot, the coefficient $a_0$ is quite
sensitive to the purity and its range is narrowing for decreasing
purity. This behaviour can be understood by the analytic form of 
coefficient $a_0$, we have 
\begin{equation*}
a_0 = \frac{1}{4} (2 + \sqrt{2 \mu -1}\ f(\alpha_1,\alpha_2,\alpha_3,\theta,\phi,\alpha,\beta))
\end{equation*}
where $f(\alpha_1,\alpha_2,\alpha_3,\theta,\phi,\alpha,\beta) \in [-2,2]$.
When $\mu =1$, $a_0 \in [0,1]$, while for $\mu = 1/2$ the only allowed
value is in fact $a_0 = 1/2$.
\par $ $ \par
\par
The distribution of the coefficients of $\Pi$ also depends on the
parameters $\alpha_1$, $\alpha_2$ and $\alpha_3$ of the unitary
operation. Looking at the central panel of Fig.\ \ref{f2:coeff}, 
medium gray points are again obtained for the free parameters randomly
chosen in their whole range, whereas light gray points are given for $\alpha_1 \in
[-\pi/3,0]$ and black ones for $\alpha_1 \in [-\pi/10,0]$. 
Notice that any constraint on $\alpha_1$ is also limiting the ranges of 
the other two parameters $\alpha_2$ and $\alpha_3$, through the
conditions given in (\ref{eq:const_alpha}).
As it is apparent from the plot
by shrinking the range of the parameter $\alpha_1$ the range
of $\vert a \vert$ is also shrinking. The limiting case is 
$\alpha_1\rightarrow 0$ (and thus $\alpha_2,\alpha_3\rightarrow 0$), 
corresponding to $\vert a \vert \rightarrow 0$ and $a_0\in [0,1]$, i.e. 
to the trivial case $V=\Id \otimes \Id$ and 
$\Pi = \Tr{P}{\rho_P\ P} \Id_S$.
\par
Consider now the case in which the range of $\alpha_2$ is narrowed up to
the point $\pi$: the constraints given for the $\alpha_i$'s force the
range of $\alpha_1$ to $-\pi$ and the range of $\alpha_3$ to 0.  This
case is again described by the central panel Fig.\ \ref{f2:coeff},
but now the light gray points are obtained taking $\alpha_1 \in
[-\pi,-3/4 \pi]$, $\alpha_2 \in [3/4 \pi, -\alpha_1]$ (notice that the
ranges of $\alpha_1$ and $\alpha_2$ are chosen in order to always keep 
$\alpha_3 \leq 0$). The black points now correspond to $\alpha_1
\in [-\pi,-9/10 \pi]$ and $\alpha_2 \in [9/10 \pi, -\alpha_1]$. It is
worth noting that, when $\alpha_1 = -\pi$, $\alpha_2 = \pi$ and
$\alpha_3 = 0$, then $V = i\, \sigma_x \otimes \sigma_x$ and $\Pi = i\,
\Tr{P}{\rho_P\ \sigma_x P \sigma_x} \Id_S$.  
\par
Let us now consider the right panel Fig.\ \ref{f2:coeff}. Here, we analyze the
distribution of the coefficients $\{a_0,\vert a \vert\}$ when the range
of $\alpha_3$ is narrowed to the point $-\pi/2$. Due to the constraints,
we have that also the ranges of $\alpha_1$ and $\alpha_2$ tend to a single 
point $\alpha_1=-\pi$ and $\alpha_2=0$.  The light gray points
correspond to $\alpha_1 \in [-\pi,-3/4 \pi]$, $\alpha_2 \in [0,-\alpha_1/3]$ and
$\alpha_3 \in [(\alpha_1 + \alpha_2)/2, -\pi/6]$, whereas black points are
for $\alpha_1 \in [-\pi,-9/10 \pi]$, $\alpha_2 \in
[0,-\alpha_1/9]$ and $\alpha_3 \in [(\alpha_1 + \alpha_2)/2, -\pi/3]$.
It is clear that the distribution of the coefficients tends to shrink to
the region close to $a_0 = 1/2$ and $\vert a \vert = 1/2$. We remark that an
effect with $a_0 = \vert a \vert = 1/2$ is a projector. In fact, for
$\alpha_1 = -\pi$, $\alpha_2 = 0$ and $\alpha_3 = -\pi/2$, the operator
$V$ is the swap operator and $\Pi$ reduces to 
$\ketbra{\xi}{\xi}$.
\par
Finally, we mention that the parameters 
$\theta$, $\phi$, $\alpha$ and $\beta$ modify the range of
the coefficients $a_i$'s, but their changes do not influence neither
$a_0$ nor $\vert a \vert$.
\section{Tradeoff between information and disturbance}\label{s3}
If we perform the measurement of an observable on a system prepared in 
a state which is not an eigenstate of the measured observable, the 
post-measurement state is different from the initial state of the
system, i.e. the system has been {\em disturbed}. At the same time, 
the outcome of the measurement provides some amount of {\em information} 
about the state of the system under investigation before the
measurement. A question thus arises on whether one may quantify 
the overall information that can be extract from a measurement as 
well as the disturbance introduced by the same
measurement\cite{ban,mg,mis,jmir,mg1,mg2,so3,ban06}.
\par
Consider the case in which a system in a generic pure state $\ket{\psi}$ undergoes
a measurement described by a POVM, composed by the effects $E_k$'s. The post-measurement
state conditioned on the occurrence of the outcome $k$ is given by
\begin{equation*}
\ket{\psi_k} = \frac{\sqrt{E_k}}{\sqrt{p_k}} \ket{\psi}
\end{equation*}
where $p_k$ is the probability distribution of the outcomes $k$'s 
for the state $\ket{\psi}$. Therefore, the disturbance introduced 
from the measurement is given by the fidelity of disturbance
$
F = \int\! d\psi\, \sum_k\, p_k\, \vert \langle{\psi_k}|{\psi}\rangle \vert^2
$
where the integral is made on all the possible initial state (e.g.\ for qubit, giving a
parametrization on the Bloch sphere, we have $d\psi = d\theta d\phi \sin \theta$). 
Notice that, if $F$ is equal to 1, then the measurement is not disturbing the system.
When the outcome of the measurement is $k$, we may infer that the 
initial state was $\ket{\phi_k}$, where $\{|\phi_h\rangle\}$ is an
arbitrary set of states. Therefore, measuring the observable, we obtain some information.
The gained information is given by the fidelity of information
$
G = \int\! d\psi\, \sum_k\, p_k\, \vert \langle{\psi}|{\phi_k}\rangle \vert^2
$.
For the qubit POVM of Eq. (\ref{effect}) the above expressions 
reduce to 
\begin{align*}
F &= \frac{1}{6} \left( 2 + 
\left|\tr{\sqrt{\Pi}}\, \right|^2 +
\left|\tr{\sqrt{\Id-\Pi}}\, \right|^2
\right)\\
G &= \frac{1}{6} \Bigg( 2 + 
\langle{\phi_0}|\Pi|\phi_0\rangle+
\langle{\phi_1}|\Id-\Pi|\phi_1\rangle
\Bigg)
\end{align*}
The ostensible freedom in the choice of the set of states 
$\ket{\phi_k}$'s is removed by maximizing the fidelity of information 
$G$. Then, each state $\ket{\phi_k}$ has to be the eigenstate of the 
effect $E_k$ with the maximum eigenvalue.
Upon exploiting Eq. (\ref{effect}), one may show that $F$ and $G$ have 
to satisfy the following relation\cite{ban}
\begin{equation}\label{td}
(F - \frac{2}{3})^2 + 4 (G - \frac{1}{2})^2 \leq \frac{1}{9}\,
\end{equation} 
which expresses quantitatively the tradeoff between information and 
disturbance in quantum measurement on a qubit. A POVM leading
to fidelities $F$ and $G$ saturating the above inequality is said to
be optimal.
\begin{figure}[h]
\includegraphics[width=0.24\textwidth]{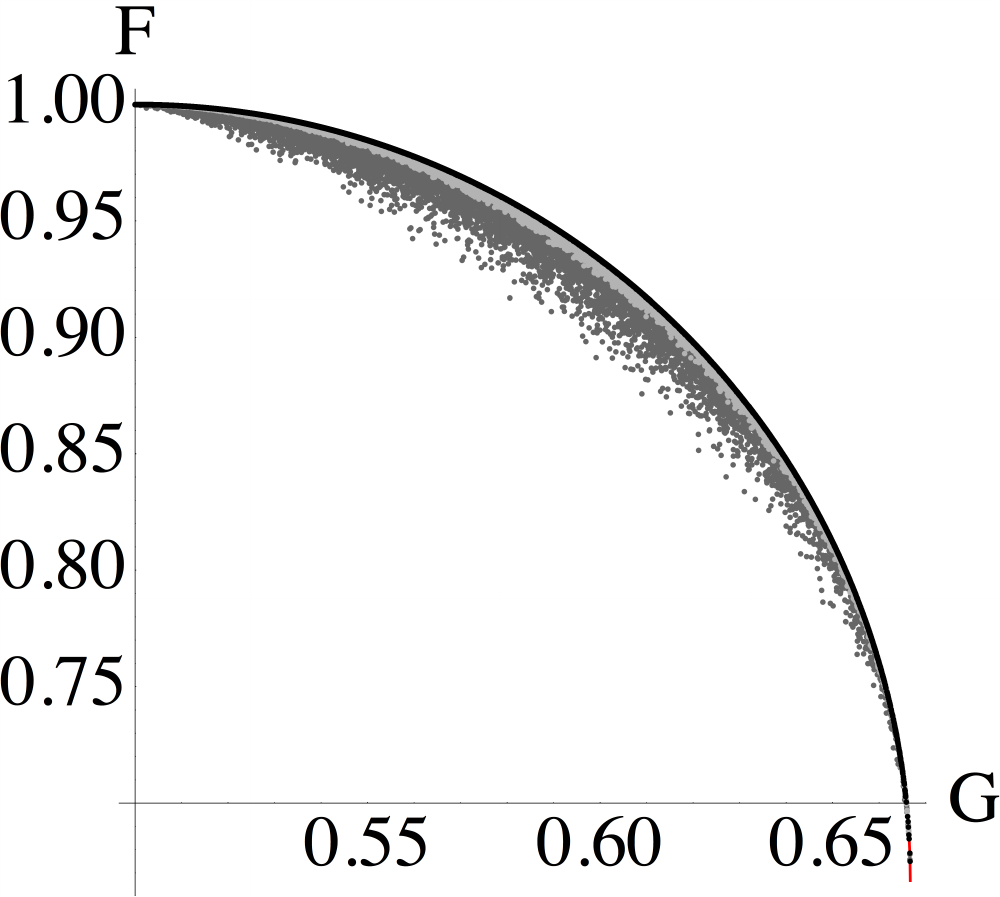}
\includegraphics[width=0.24\textwidth]{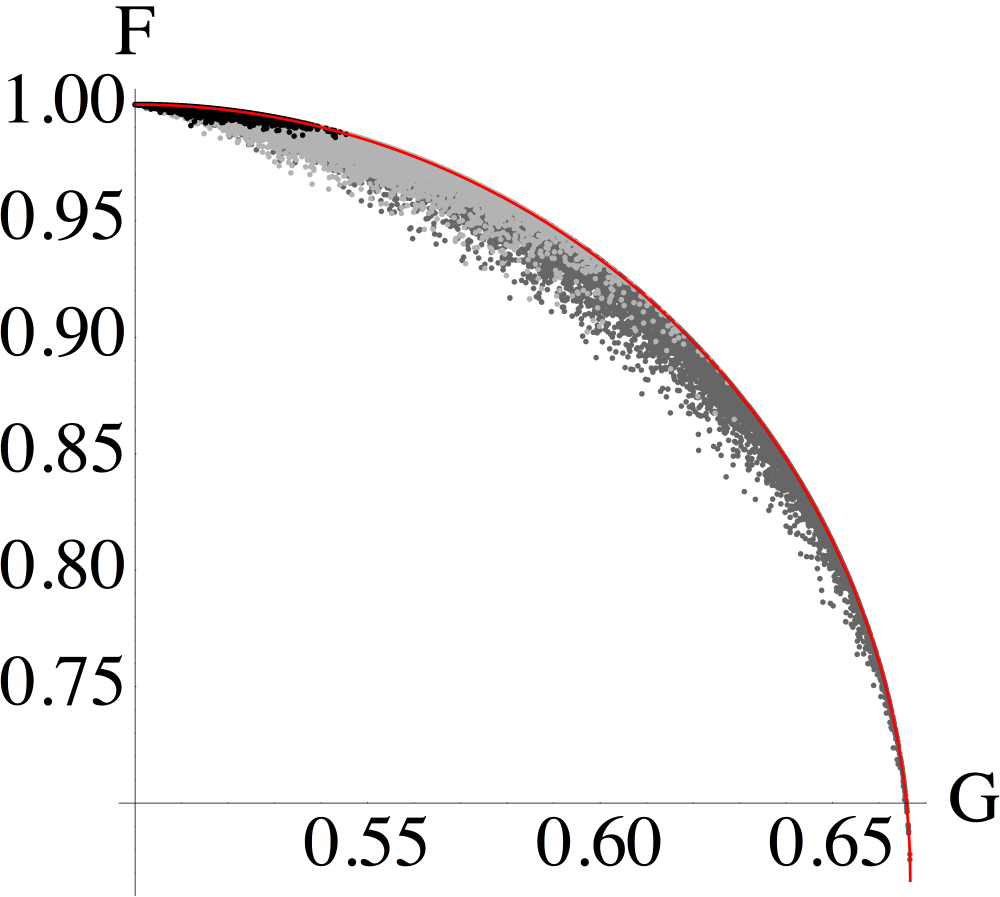}
\includegraphics[width=0.24\textwidth]{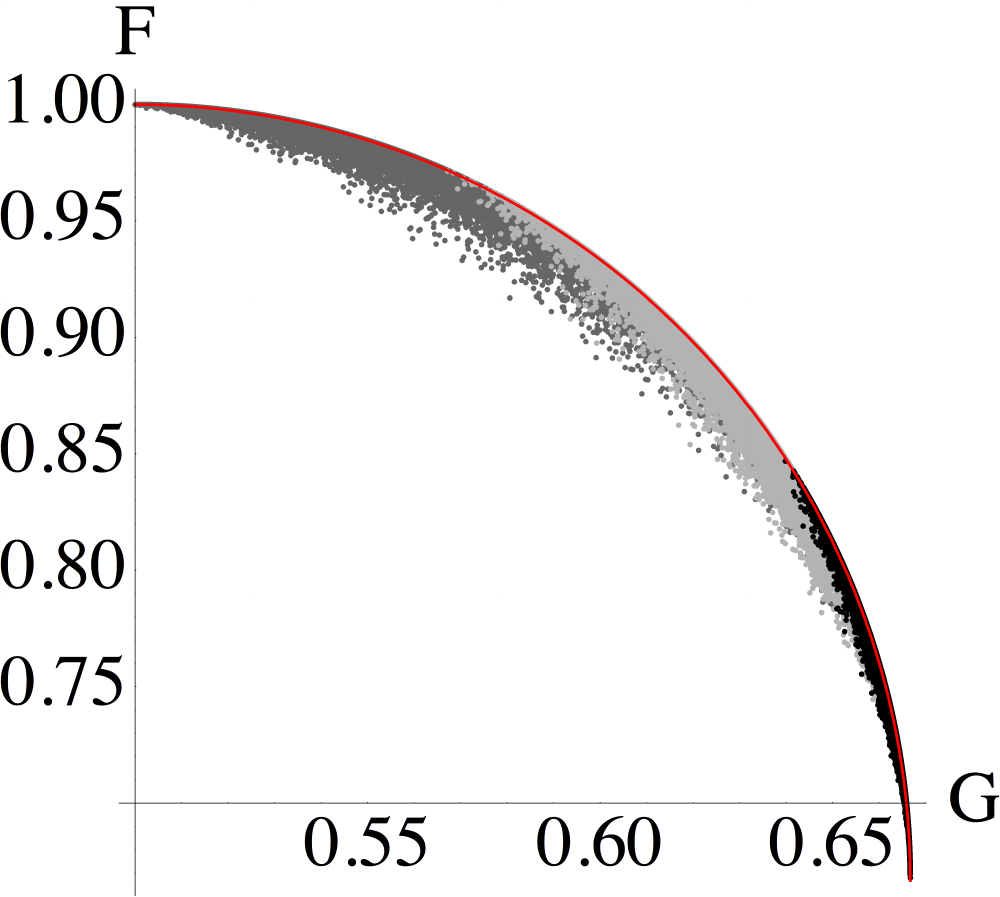}
\includegraphics[width=0.24\textwidth]{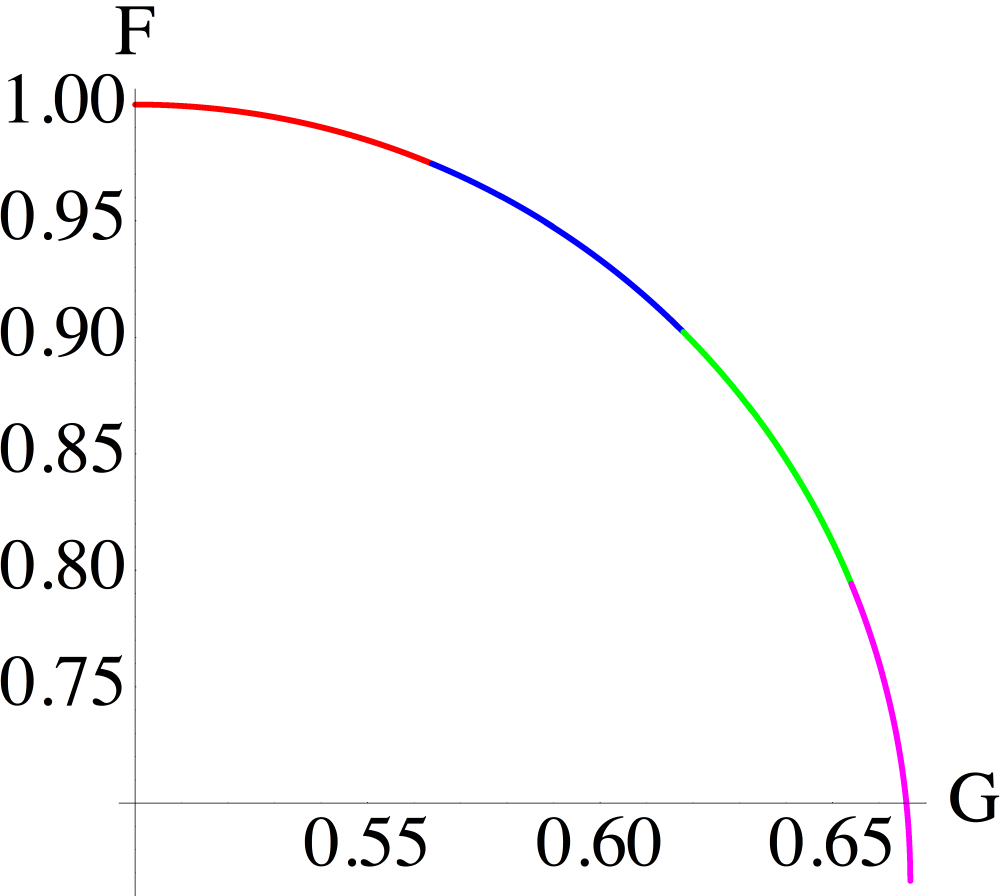}
\caption{(Color online) Tradeoff between the information fidelity 
$G$ and the disturbance fidelity $F$ for different ranges of the free parameters.
In the three panels on the left, the medium gray points correspond
to POVM obtained by choosing the free parameters in their 
whole range, whereas the light gray points and
the black ones to POVMs for restricted ranges of some parameters.
The solid red line denotes the optimal tradeoff, i.e. the values
saturating the inequality in Eq. (\ref{td}).
In the most left panel, we show the distribution of $G$ and $F$ for
different ranges of the probe purity:
light gray points correspond to $\mu \in [0.5,0.7]$ and the 
black ones to $\mu \in [0.5,0.51]$. The second plot shows results
for different ranges of $\alpha_1$: the light gray points corresponds 
to $\alpha_1 \in [-\pi/3,0]$ and the black ones for 
$\alpha_1 \in [-\pi/10,0]$. 
The same distributions are obtained for $\alpha_1 \in [-\pi,-3/4 \pi]$,
and $\alpha_2 \in [3/4 \pi, -\alpha_1]$ (light gray points), and 
$\alpha_1 \in [-\pi,-9/10 \pi]$ and $\alpha_2 \in [9/10 \pi, -\alpha_1]$
(black points). In the third panel the light gray points corresponds to
$\alpha_1 \in [-\pi,-3/4 \pi]$, $\alpha_2 \in [0,-\alpha_1/3]$ and 
$\alpha_3 \in [(\alpha_1 + \alpha_2)/2, -\pi/6]$, whereas black points are 
for $\alpha_1 \in [-\pi,-9/10 \pi]$,
$\alpha_2 \in [0,-\alpha_1/9]$ and 
$\alpha_3 \in [(\alpha_1 + \alpha_2)/2, -\pi/3]$.
The last panel on the right shows the fidelities obtained using 
a $C_{not}$ gate to couple signal and probe, as a function of 
the population parameter $\theta$ of the probe.}
\label{f3:tradeoff}
\end{figure}
\par
In order to understand whether there is some typical value of the
tradeoff we have performed a study of the distribution of the 
pairs $\{ G , F \}$ for POVMs obtained for different distributions 
of the free parameters. In particular, we have considered the same 
ranges used in the previous Section for $\mu$ and the $\alpha_k$'s.
The results are shown in Fig.\ \ref{f3:tradeoff}, where again, 
the medium gray points are obtained
by taking at random the free parameters into their whole range 
of variation.
\par
\begin{figure}[h]
\includegraphics[width=0.32\textwidth]{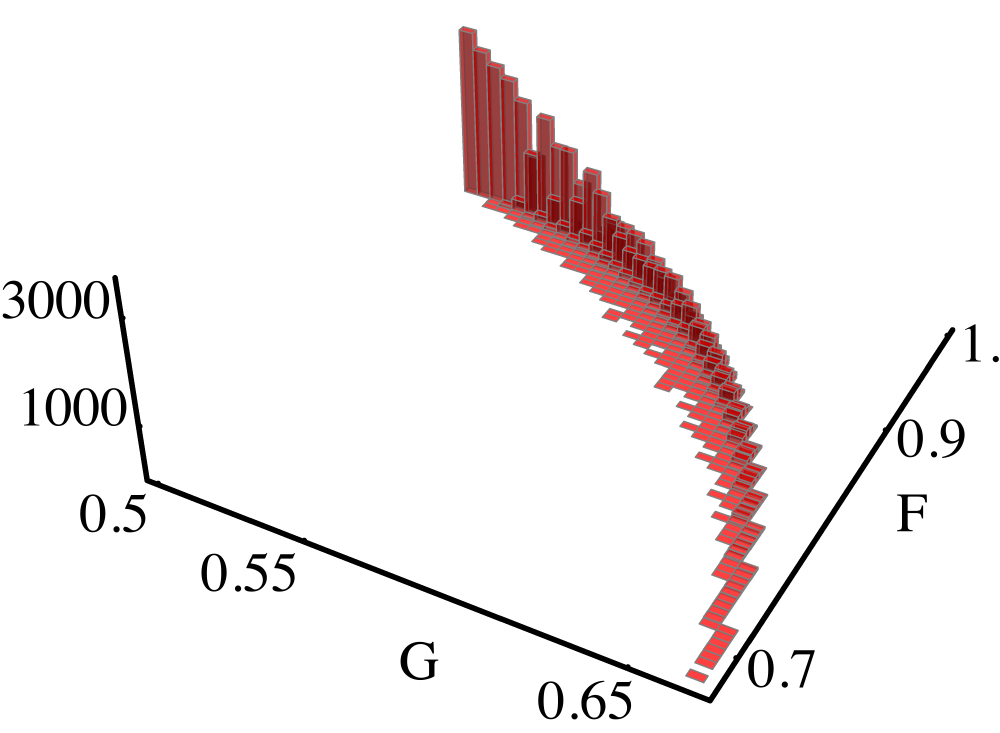}
\includegraphics[width=0.32\textwidth]{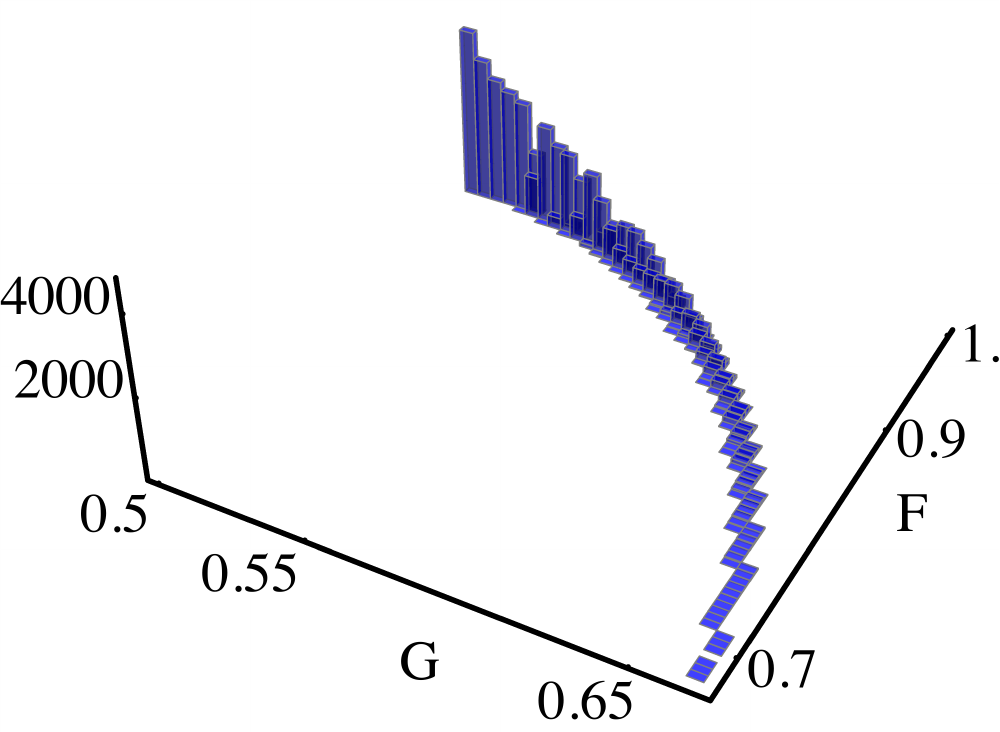}
\includegraphics[width=0.32\textwidth]{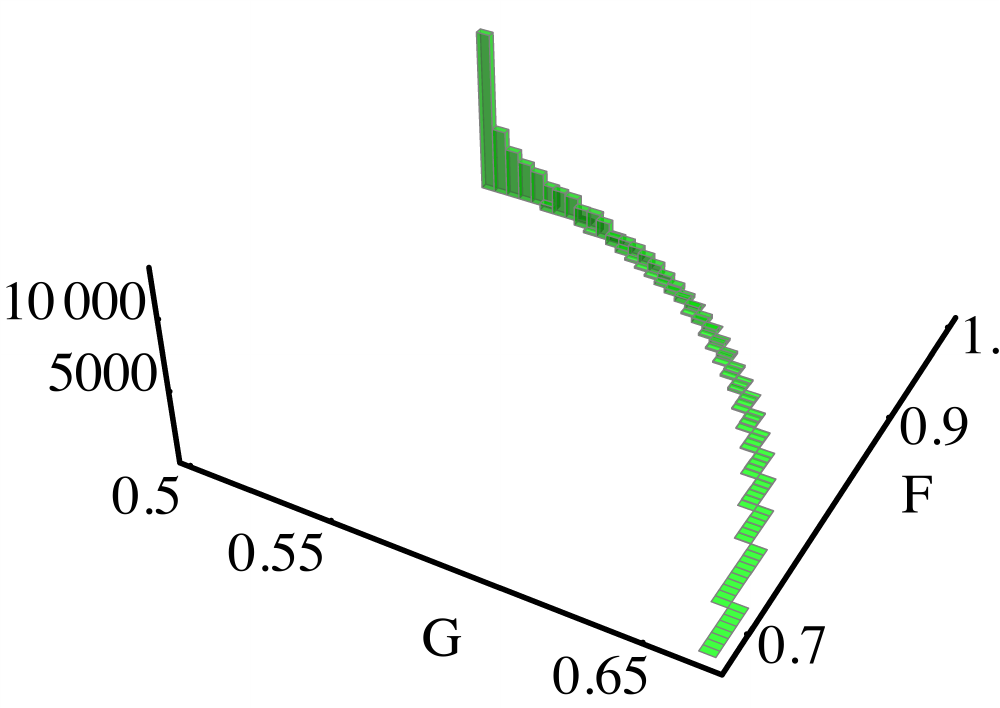}
\caption{(Color online) The distribution of the POVMs as a function of
$G$ and $F$. The first histogram is obtained by taking $\mu$ into its
whole range. The second one for $\mu \in [0.5,0.75]$. The last one for
$\mu = 1/2$.}
\label{f4:isto}
\end{figure}

In the left panel of Fig.\ \ref{f3:tradeoff}, the distribution of $\{ G
, F\}$ is shown for different ranges of the purity $\mu$ of the probe 
system.  In particular, light gray points are taken
for $\mu \in [0.5,0.7]$, while the black ones are taken for $\mu \in
[0.5,0.51]$. As it is apparent from the plot, by narrowing 
the range of $\mu$ the resulting POVMs become closer and closer
to the optimal ones. In the limiting case of $\mu=\frac12$ all the
resulting POVMs have a tradeoff falling on the optimal curve of 
Eq. (\ref{td}), i.e. all the POVMs are optimal. In order to
have a more detailed picture, the histograms of their distribution are 
shown in Fig. \ref{f4:isto}: in the left panel, the POVMs are
generated by choosing at random the parameters into their whole ranges.
The histogram displays a distribution with a maximal value at the point
$G = 1/2$ and $F = 1$, i.e. POVMs that neither gain information, nor
disturb the state of the system.  Moreover, it is apparent that not all
the produced POVMs are optimal. The second histogram is obtained by
taking at random $\mu$ between $0.5$ and $0.75$; in this case the
distribution is different from zero for values of $F$ and $G$ near the
optimal limit. In the right histogram, the distribution is taken
for $\mu = 1/2$: all the POVMs are optimal, but the distribution has a
peak at the point $G=1/2$, $F=1$. Overall, 
the emerging picture is that even using a maximally mixed probe it is
possible to saturate the optimal tradeoff. On the other hand, in this
case the {\em typical} POVM is the non-informative one $\Pi=\Id$. Still, 
it is possible to find POVMs with $G=F=2/3$, that is a measurement which
extracts maximal information from the system and introduces a maximal
disturbance.  The two-qubit operator that gives this kind of POVMs is
the swap operator $V(-\pi,0,-\pi/2)$.  
\par
In the other panels of Fig.\ \ref{f3:tradeoff}, we show how the
distribution of the fidelities $\{ G,F \}$ is affected by the ranges of 
$\alpha_1$, $\alpha_2$ and $\alpha_3$.
In particular, the second panel refers to
the case in which the range of the parameter $\alpha_1$ is progressively
shrinking to the single point $\alpha_1=0$.
As previously said, the constraints on the parameters $\alpha_i$'s force 
the other two parameters $\alpha_2$ and $\alpha_3$ to narrow their 
ranges into the single point $\alpha_2=\alpha_3=0$. 
The light gray points are taken for $\alpha_1 \in [-\pi/3,0]$ and
black ones for $\alpha_1 \in [-\pi/10,0]$. 
The behavior of the distribution is quite clear: it collapses
into the point $G=1/2$ and $F=1$, when $\alpha_i$'s $\rightarrow 0$.
We recall that the corresponding POVMs are 
proportional to the identity $\Id_S$.
\par
The second panel of Fig.\ \ref{f3:tradeoff} also describes the trend of
the distribution when the range of $\alpha_2$ is narrowed to the point
$\pi$. Again, the constraints force the range of $\alpha_1$ to $-\pi$
and the range of $\alpha_3$ to 0.  In this case, the light gray points
are obtained taking $\alpha_1 \in [-\pi,-3/4 \pi]$, $\alpha_2 \in [3/4
\pi, -\alpha_1]$, while the black ones are taken for $\alpha_1 \in
[-\pi,-9/10 \pi]$ and $\alpha_2 \in [9/10 \pi, -\alpha_1]$.  
\par
The third panel of Fig.\ \ref{f3:tradeoff} refers to the case in
which the range of $\alpha_3$ is gradually reduced to the point
$-\pi/2$, and therefore $\alpha_1\rightarrow-\pi$ and
$\alpha_2\rightarrow 0$. 
The light gray points are taken for $\alpha_1 \in [-\pi,-3/4 \pi]$,
$\alpha_2 \in [0,-\alpha_1/3]$ and $\alpha_3 \in [(\alpha_1 +
\alpha_2)/2, -\pi/6]$, 
while black points stay for $\alpha_1 \in [-\pi,-9/10 \pi]$,
$\alpha_2 \in [0,-\alpha_1/9]$ and $\alpha_3 \in [(\alpha_1 + \alpha_2)/2, -\pi/3]$.
The distribution of $\{ G , F \}$ collapses into the point
$F = G = 2/3$. In fact, for $\alpha_1 = - \pi$, $\alpha_2 = 0$ and $\alpha_3 = -\pi/2$,
we obtain a projective measurement, giving as more information as 
possible about the system, at the price of introducing a considerable
disturbance.
\par
Finally, in the right panel of Fig. \ref{f3:tradeoff}, we show the
fidelities obtained by using a $C_{not}$ gate to couple the system and
the probe qubit and then measuring $\sigma_3$ on the probe.
In particular, we have considered the fidelities obtained by 
varying the $\theta$ parameter of the probe: 
the red portion of the curve corresponds to $\theta \in
[0,\pi/8]$, the blue one to $\theta \in [\pi/8,\pi/4]$, green is for 
$\theta \in [\pi/4,3/8 \pi]$, and magenta for $\theta \in [3/8\pi,\pi/2]$.
As it is apparent from the plot, we confirm the known 
optimality\cite{mg,mis} of  the resulting POVMs.
Notice that, since the Cartan decomposition has
been used to obtain the coefficients of the effect $\Pi$, we need to
find the operator $V(\alpha_1,\alpha_2,\alpha_3)$ connected to the
$C_{not}$ gate.  After straightforward calculation we find that
$C_{not} =  ( R_1 \otimes R_2 )\
V(-\frac{\pi}{2},\frac{\pi}{2},0)\ ( S_1 \otimes S_2 )$
where, as shown before, the local operators do not modify neither $F$
nor $G$.  
\section{Conclusions}\label{con}
In this paper, we have addressed the properties of the class of 
two-value qubit POVMs $\{\Pi,\Id-\Pi\}$ that are obtained by coupling 
the signal qubit with a probe qubit and then performing a projective 
measurement on the sole probe system. 
These POVMs represent the simplest class of qubit POVMs and 
depends on $3+3+2=8$ free parameters describing the initial preparation of the 
probe qubit, the Cartan representative of the unitary coupling, and the
projective measurement at the output, respectively. 
We have obtained the analytic expression of the coefficients
$(a_0, a_1, a_2, a_3)$ of the effect $\Pi$ in the Pauli basis
and have used these expressions to understand which parameters
are relevant to specific properties of the POVMs. 
In particular, for  the distribution of $\{a_0,\vert a \vert\}$ 
we found that the relevant parameters are the purity $\mu$ of the 
probe system and the parameters defining the Cartan representative
of the unitary coupling.
We have also analyzed in details 
the tradeoff between information and disturbance for different ranges of the 
free parameters, showing, among other things, that i) typical values of the 
tradeoff are close to optimality and ii) even using a maximally mixed probe 
one may achieve optimal tradeoff (using a swap gate to couple the signal
and the probe qubit), though the typical POVM is the non-informative one $\Pi=\Id$. 
\section*{Acknowledgments}
This work has been supported by MIUR through the project 
FIRB-RBFR10YQ3H-LiCHIS.
\appendix
\section{The matrix elements the effect $\Pi$ in the Pauli basis}\label{a:cote}
The effect $\Pi = a_0\ \Id + \boldsymbol{a} \cdot \boldsymbol{\sigma}$ 
has the following coefficient:
\begin{align*}
a_0 = &\frac{1}{4} (2+\sqrt{2 \mu-1}\, (\cos \alpha \cos \theta\, (\cos \alpha_1+\cos \alpha_2)+2 \cos \alpha_3 \sin \alpha \sin \theta\\
&(\cos(\frac{\alpha_1+\alpha_2}{2}) \cos \beta \cos \phi+\cos(\frac{\alpha_1-\alpha_2}{2}) \sin \beta \sin \phi)))\\
a_1 = &\frac{1}{4} (2 \cos \beta \sin \alpha \sin(\frac{\alpha_1+\alpha_2}{2}) \sin \alpha_3\\
&+\sqrt{2 \mu-1}\, (\cos \alpha\, (\sin \alpha_1-\sin \alpha_2) \sin \theta \sin \phi-2 \cos \alpha_3
\cos \theta \sin\alpha \sin (\frac{\alpha_1-\alpha_2}{2}) \sin \beta))\\
a_2 = &\frac{1}{4} (2 \sin \alpha \sin (\frac{\alpha_1-\alpha_2}{2}) \sin \alpha_3 \sin \beta\\
&+\sqrt{2 \mu-1}\, (2 \cos \alpha_3 \cos \beta \cos \theta \sin \alpha \sin (\frac{\alpha_1+\alpha_2}{2})
-\cos \alpha \cos \phi\, (\sin \alpha_1+\sin \alpha_2) \sin \theta))\\
a_3 = &\frac{1}{4} (\cos \alpha\, (\cos \alpha_2-\cos \alpha_1)\\
&+2 \sqrt{2 \mu-1} \sin \alpha \sin \alpha_3 \sin \theta\, (\cos (\frac{\alpha_1-\alpha_2}{2}) \cos \phi \sin \beta
-\cos (\frac{\alpha_1+\alpha_2}{2}) \cos \beta \sin \phi))\\
\end{align*}

\end{document}